# Efficient Random Walk based Sampling with Inverse Degree

Xiao Qi

x.qi274@gmail.com

## Abstract

Random walk sampling methods have been widely used in graph sampling in recent years, while it has bias towards higher degree nodes in the sample. To overcome this deficiency, classical methods such as MHRW design weighted walking by repeating low-degree nodes while rejecting high-degree nodes, so that the long-term behavior of Markov chain can achieve uniform distribution. This modification, however, may make the sampler stay in the same node for several times, leading to undersampling. To address this issue, we propose a sampling framework that only need current and candidate node degree to improve the performance of graph sampling methods. We also extend our original idea to a more general framework. Our extended IDRW method finds a balance between the large deviation problem of SRW and sample rejection problem in MHRW. We evaluate our technique in simulation by running extensive experiments on various real-world datasets, and the result show that our method improves the accuracy compared with the state of art techniques. We also investigate the effect of the parameter and give the suggested range for a better usage in application.

# 1. Introduction

Over the past two decades, network science has demonstrated its strength in modeling many real-world interactive systems that are generic agents connected by pairs of edges, i.e., nodes. Such models have applications in many areas, such as social networks (Backstrom et al., 2006; Cem and Sarac, 2016, 2016), transportation networks (Duan and Lu, 2014; Xie and Levinson, 2007), collaboration networks (Newman, 2001) etc. However, exploring such systems becomes very difficult due to the huge number of users and relationships in real-world systems. Therefore, graph sampling becomes an important technique, scaling down the size of original graphs while capturing the key properties of the large-scale graphs.

For different application scenarios, scientists have adopted different approaches for different application scenarios. The most ideal sampling method is to directly sample the edges/nodes uniformly, which is called Uniform Node Sampling. However, due to the privacy protection policies of most online social networks and the high price of uniform sampling, most of the existing sampling methods crawl nodes/edges to form samples by random walk. The obtained sample data is then estimated according to the random walk method, thus the properties of our graph of interest would be preserved.

There are two principles in graph sampling, the first is to collect as uniform a sample as possible; the second is that a faster converging random walk process is better in a limited sample size. So, for both principles, different algorithms have emerged to target improvements. To get a uniform distribution as the stationary distribution of random walk, Metropolis-Hastings technique (Hastings, 1970) is implemented into a simple random walk process, with uniform distribution as the target distribution (Thompson, 2006). That is the widely used algorithm called MHRW. MD (Bar-Yossef and Gurevich, 2008) was proposed to revise the topology of graphs so that we can get a uniform distribution of node by running a simple random walk on the revised graph. Although their stationary distribution is the most ideal one, they have deficiencies. MHRW has problems of rejecting samples, which leads to a small actual sample size in the estimator; MD and the improved version GMD (Li et al., 2015a) also has a problem of repeating problems, and a unnoticeable problem of getting stuck in low-degree nodes by self-loops.

Researchers also proposed new sampling methods by the second principle – overcoming the slow mixing problem (Li et al., 2019). In many social media platforms, the numbers of the follower and the followees are displayed directly, while the list of followers and followees need to send further request. So, in social platforms such as Twitter, sampling algorithms demanding information about neighbors, such as Common Neighbor Awareness Random Walk (CNARW) are not feasible if we can only get the information about node degrees. Therefore, making use of easy-to-reach information to overcome the problem of previous techniques in sampling OSN is the research gap we will fill in this paper.

In this paper, we begin with a simple idea of using the inverse of the degree to filter whether to accept the candidate node. this is our first model Inverse degree random walk (IDRW). After an investigation of the mathematical properties, we extend the method to a more general framework, containing α as a parameter, so that our sampling framework is more flexible by parameter adjustment in the real-world application. We also conducted extensive experiments on real-world datasets, investigating the effect of query cost on error and the effect of α on error. Based on the results, we also give application recommendation.

The reminder of this paper will proceed as follows. In Section 2, we provide preliminaries about graph sampling. In Section 3 and Section 4, we elaborate our original idea and extend the method to a more general framework, respectively. In Section 5 we conduct experiments to validate our methods and discuss the pros and cons, and Section 6 concludes.

## 2. Preliminaries and Notations

In this section we introduce the basic notation of graphs and preliminaries needed in random walk on graphs. We give descriptions of random walk, variants of RW, and their theoretical analysis.

### 2.1 Random walk on graphs

If we have an undirected and connected graph that can be denoted by $G\ (V, E)$, where $V$ is the node set and $E$ is the edge set. As commonly used statistics of graphs, the number of node and edges of $G$ are denoted by $|V|$ and $|E|$ ,respectively. For convenience, we also define neighbor set of a node $i$ as $N(i)$, and the degree of $i$, i.e., the number of neighbors of $i$, is denoted as $d_i$. From the diffusion aspect, Random Walk (RW) on graph $G$ is a process that starts from one node, say $v_0 \in V$, and randomly chooses a neighbor of the current node as the next node. Moving to the nodes step by step makes up of a complete random walk on a graph. Random walk can be comprehended in an abstract way; it can be viewed as a discrete- time Markov chain in finite state space. The state space contains all nodes in the node set $V$, and the edges, which link different states of the Markov chain, represents different probability in different selection mechanisms.

Here we introduce Simple Random Walk (SRW), which is the foundation of random walk algorithms. The selection mechanism is uniformly choosing a neighbor of current node $i$, so in SRW, the probability of moving from $i$ to $j$ is

$$P(X_{t+1} = j | X_t = i) = P_{ij} = \begin{cases} \frac{1}{d_i}, & if\ j \in N(i), \\ 0, & otherwise. \end{cases}$$

Note that the graph is not equivalent to the Markov chain diagram. For an undirected and connected graph $G$, walking from $i$ to $j$ depends on $d_i$ while walking from $j$ to $i$ depends on $d_j$. The adjacent matrix of a graph $G$ is denoted by $A$.

$$A = \begin{pmatrix} a_{11} & \cdots & a_{1N} \\ \vdots & \ddots & \vdots \\ a_{N1} & \cdots & a_{NN} \end{pmatrix}$$

thus, the walk matrix, also the transition matrix of SRW is

$$P = \begin{pmatrix} a_{11}/\sum a_{1\cdot} & \cdots & a_{1N}/\sum a_{1\cdot} \\ \vdots & \ddots & \vdots \\ a_{N1}/\sum a_{N\cdot} & \cdots & a_{NN}/\sum a_{N\cdot} \end{pmatrix}$$

where $\sum a_{k\cdot}$ represents the sum of all elements in $k$th row in $A$, and $p_{ij}$ in $P$ represents the probability that random walker can walk from node $i$ to node $j$ in one step. Therefore, we can make use Markov chain theory to get the stationary distribution of the process. Given an initial distribution $\pi_0$, which represents the initial proportion of visited nodes, we use the detailed balance condition

$$\pi_i P_{ij} = \pi_j P_{ji}$$

and $\pi$ is a probability density function, implying $\sum_{i=1}^{N} \pi_i = 1$, to calculate the stationary distribution of the Markov chain. Finally, we have the stationary distribution of the Markov chain

$$\pi_i = \frac{d_i}{\sum d_i} = \frac{d_i}{2|E|}.$$

**2.2 Unbiased estimator**

Now that we have a sample collected by SRW, we need an estimator with good mathematical properties. Therefore, we can derive that if we employ SRW as the sampling method to collect nodes, the probability of being selected of a node is proportional to its degree. So, in the estimation framework, researchers usually use Horvitz–Thompson estimator (Horvitz and

Thompson, 1952) to derive an asymptotical unbiased estimator. When random walker reaches the steady state, a node $i$ is sampled with probability $\pi(i)$, where $\pi(i)$ differs according to the stationary distributions generated by neighbor selection mechanism. If we denote the set of sampled nodes as $S_0$, an asymptotically unbiased estimator of $\theta$ is given by

$$\hat{\theta} = \frac{\sum_{i \in S_0} \frac{f(i)}{\pi(i)}}{\sum_{i \in S_0} \frac{1}{\pi(i)}} \tag{1}$$

The theory backing up the unbiasedness of estimator is the Law of Large Numbers (LNN) of Markov chains.

**Theorem 1**: (Law of Large Number). Let $S$ be a sample path obtained by a Markov chain defined on a state space $V$ with stationary distribution $\pi$. For any function $f, g: V \mapsto \mathbb{R}$, and let $F_s(f) \triangleq \sum_{i \in V} \pi_i f(i)$. It holds that

$$\lim_{|S| \to \infty} \frac{1}{|S|} F_S(f) = E_\pi[f] \tag{2}$$

and

$$\lim_{|S| \to \infty} \frac{F_S(f)}{F_S(g)} = \frac{E_\pi[f]}{E_\pi[g]} \tag{3}$$

almost sure. Now that we can make use LNN, the estimated value $\hat{\theta} = \frac{\sum(f(i)/\pi(i))}{\sum(1/\pi(i))}$ converges to $\theta$. So, estimator (1) is an asymptotically unbiased estimator.

## 3. IDRW: Inverse Degree Random Walk

In Section 3, we present the details of our inverse degree random walk. We first introduce the main idea of our new method by using s simple graph. Then we present the design of transition matrix of our method. We also provide mathematical analysis of stationary distribution to make sure that an asymptotically unbiased estimator can be derived from the sample collected by IDRW.

### 3.1 Motivation and main idea

SRW on connected graphs usually have problems of large deviation (Li et al., 2015a), and the deviation is proved to be dependent on the similarity between the stationary distribution of SRW (proportional to the node degree) and uniform distribution. In many real world networks, their degree distribution follows power law (Adamic et al., 2008), i.e., the degrees of nodes in the same graph may vary from 1 to $10^3$, so the stationary has a large deviation, leading to a poor estimation. Therefore, one direction to improve SRW algorithm is to reweigh the edge by modifying the selection strategy in every step, thus the probability of being reached by random walker would be adjusted to a stationary distribution with smaller deviation.

Intuitively, the random walk will travel all edges given long enough time with same frequencies, so the edge distribution is uniform, and every node can be reached by the edges connecting them. Therefore, node $i$ has a probability of being reached with $\frac{d_i}{2|E|}$. In SRW, every edge connecting current node has a probability to be visited $1/d_i$, it only depends on the current node information, the candidate nodes (neighbors) degrees do not participate in the decision. In our newly proposed methods, our mean idea is to control the edges being visited by both degrees of current node and the candidate node. In the selection procedure, the random walker will choose a neighbor uniformly and use the inverse of candidate node degree to filter the easy-to-reach nodes. We use a simple graph in Fig.1 to get a clearer illustration.

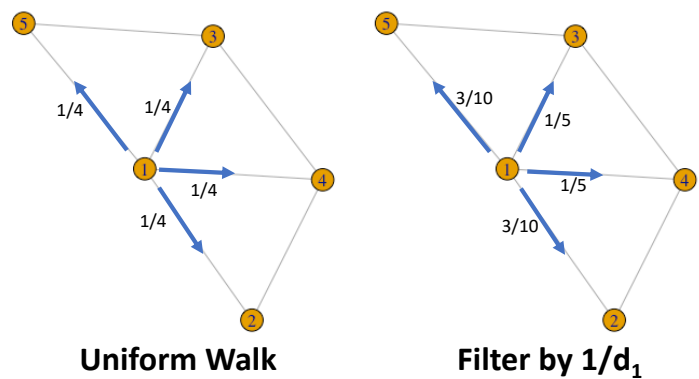

Figure 1 Different Walk on Simple Graph G

In Fig. 1, we can write the adjacent matrix of the graph

$$A = \begin{pmatrix} 0 & 1 & 1 & 1 & 1 \\ 1 & 0 & 0 & 1 & 0 \\ 1 & 0 & 0 & 1 & 1 \\ 1 & 1 & 1 & 0 & 0 \\ 1 & 0 & 1 & 0 & 0 \end{pmatrix}$$

thus, the walk matrix, also the one-step transition matrix of SRW is

$$P = \begin{pmatrix} 0 & \frac{1}{4} & \frac{1}{4} & \frac{1}{4} & \frac{1}{4} \\ \frac{1}{2} & 0 & 0 & \frac{1}{2} & 0 \\ \frac{1}{3} & 0 & 0 & \frac{1}{3} & \frac{1}{3} \\ \frac{1}{3} & \frac{1}{3} & \frac{1}{3} & 0 & 0 \\ \frac{1}{2} & 0 & \frac{1}{2} & 0 & 0 \end{pmatrix}$$

But for our method, the one-step transition matrix is

$$P_{IDRW} = \begin{pmatrix} 0 & \frac{3}{10} & \frac{1}{5} & \frac{1}{5} & \frac{3}{10} \\ \frac{3}{7} & 0 & 0 & \frac{4}{7} & 0 \\ \frac{3}{13} & 0 & 0 & \frac{4}{13} & \frac{6}{13} \\ \frac{3}{13} & \frac{6}{13} & \frac{4}{13} & 0 & 0 \\ \frac{3}{7} & 0 & \frac{4}{7} & 0 & 0 \end{pmatrix}$$

For node 1, if we use SRW to choose a neighbor as the next node, $N(1) = \{2,3,4,5\}$, every node has the same probability to be chosen. In $N(1)$, node 3 and 4 are with relatively high degree while 2 and 5 have a lower degree than 3 and 4. In our method, we reassign the probability of being chosen based on their node degree. The nodes with lower degree (2 and 5) have larger probabilities of being chosen while the nodes with higher degree (3 and 4) have smaller probabilities of being chosen. It is clear that our method can correct the bias towards high degree nodes of SRW, and lead to a stationary distribution with smaller deviation than SRW.

### 3.2 Design of Inverse Degree algorithm

To realize Inverse Degree Random Walk (IDRW) described above by using the weighted walking strategy, the key issue is to formulate the selection of the next node with a mathematical model. We bring up the transition probability matrix and the stationary distribution according to our formulation. In the following, the first step is to formulate the selection mechanism, then present the design of a transition matrix, and show the Inverse Degree Random Walk method in details. As we choose a neighbor of $v$ uniformly, the nodes connected with $i$ has the probability $1/d_i$ to be chosen. In the next step, denoting the node we have chosen as $j$, $j$ is the candidate node. To decide whether accept $j$ or not, we would compare a random number which is uniformly distributed in $(0,1)$ and the inverse degree of node $j$, i.e., $1/d_j$. The candidate will be accepted if the random number is smaller than $1/d_u$, otherwise, it will be rejected. The idea behind the algorithm is simple, if the candidate node has a high degree, $1/d_j$ is small, then our selection procedure has a smaller probability to accept the high-degree node. However, if the candidate node has a low degree, the probability of being selected is relatively high.

The details in our algorithm are displayed in the Algorithm 1.

**Algorithm 1:** One step of Inverse Degree Random Walk
**Input:** current node $i$;
**Output:** next node $j$;
  **do**
      Select $j$ uniformly at random $i$'s neighbors;
      Generate a random number $q \sim U(0,1)$
  **while** $(q > 1/d_j)$
  Return $j$;

### 3.3 Analysis of stationary distribution

To get the stationary distribution of our random walker, we need to give the transition matrix first. According to the description in Section 3.2, it is easy to write the transition probability as

$$P(X_{t+1} = j | X_t = i) = P_{ij} = \begin{cases} \dfrac{1}{d_i d_j}, & if\ j \in N(i), \\ 0, & otherwise. \end{cases} \quad (4)$$

However, it has a chance of walking back to $i$, so we need to do a normalization on the transition probabilities, and the transition matrix can be written as

$$P_{ij} = \begin{cases} \dfrac{\tilde{p}_{ij}}{1 - \tilde{p}_{ii}}, & if\ v \in N(v) \\ 0, & otherwise \end{cases} \quad (5)$$

where $\tilde{p}_{ij}$ is defined as

$$\tilde{p}_{ij} = \begin{cases} \dfrac{1}{d_i d_j}, & if\ j \in N(i) \\ 1 - \sum_{k \in N(i)} \tilde{p}_{ik}, & if\ i = j \\ 0, & otherwise \end{cases} \quad (6)$$

We also take Fig.1 as an example to illustrate the selection mechanism for one step.

In the following Theorem 2, we would prove that the designed Markov chain IDRW has a unique stationary distribution.

**Theorem 2:** If graph $G = (V, E)$ is connected and undirected, Inverse Degree Random Walk has a unique stationary distribution on $G$.

*Proof:* For any node $i$ and $j \in N(i)$, the transition probability $P_{ij}$ is larger than 0, as the value of $1/d_i$ and $1/d_j$ are always positive in a connected graph. Thus, for any two nodes $i$ and $j$ in $G$, they are reachable from each other in finite steps for Inverse Degree Random Walk since $G$ is an undirected and connected graph. Therefore, we can conclude that the Markov chain constructed by Inverse Degree Random Walk is irreducible.

According to Markov chain theory, any irreducible Markov chain on an undirected and connected graph has a unique distribution, IDRW has a unique stationary distribution. ∎

Now that IDRW has a unique stationary distribution, we would utilize the detailed balance condition to get the stationary distribution. We give the quantitative relationship between nodes in $G$ in Theorem 2.

**Theorem 3:** After IDRW converges, for any node $i \in V$, the stationary distribution of $i$ being visited $\pi_i = Z \times (1 - \tilde{p}_{ii})$, where $Z$ is a normalization constant.

*Proof:* After the walker reaches equilibrium, the Markov chain constructed by IDRW have the time reversibility. So according to Proposition 1.1 in (Sigman, 2009), we only need to show that the following equation has a unique solution.

$$\pi_i P_{ij} = \pi_j P_{ji}$$

Based on the definition of IDRW transition matrix, we have

$$\frac{\pi_i}{\pi_j} = \frac{P_{ji}}{P_{ij}} = \frac{\frac{1}{d_i d_j}/(1 - \tilde{p}_{jj})}{\frac{1}{d_i d_j}/(1 - \tilde{p}_{ii})} = \frac{(1 - \tilde{p}_{ii})}{(1 - \tilde{p}_{jj})}.$$

As $(1 - \tilde{p}_{ii})$ and $(1 - \tilde{p}_{jj})$ are fixed for a given graph, so $\frac{\pi_i}{\pi_j}$ have a fixed ratio for a given graph. The value of $\pi_i$ is also constricted by the property of a probability density function $\sum \pi_i = 1$, so there is a unique solution for all $\pi_i$, thus for the distribution $\pi$. The stationary distribution of $i$ can be derived as

$$\pi_i = Z \times (1 - \tilde{p}_{ii})$$

where $Z$ is the normalization constant. ∎

Through Theorem 3, we have derived the stationary distribution of our method, thus we can derive an asymptotically unbiased estimator for the function of interest $\theta = f(G)$.

## 3.4. Estimation

Since the stationary distribution of nodes being visited by IDRW is not a uniform, bias correction is necessary to achieve asymptotically unbiased estimation. Following the importance sampling framework, an unbiased estimator is

$$\hat{\theta} = \frac{\sum_{i \in S_0} \frac{f(i)}{\pi(i)}}{\sum_{i \in S_0} \frac{1}{\pi(i)}}.$$

## 4. An Extensive Version of IDRW

From the simple graph example, the value of $1 - \tilde{p}_{ii}$ is less than 1, which means that the probability of staying at the current node is larger than 0. Therefore, to enlarge the probability of getting out of the current node and go on collecting new nodes, we modify our algorithm through replacing $1/d_j$ by $1/d_j^\alpha$ in the selection step, where $\alpha$ is a positive number. the details are in the Algorithm 2.

**Algorithm 2:** One step of Extended Inverse Degree Random Walk (Extended IDRW)
**Input:** current node $i$;
**Output:** next node $j$;
  **do**
    Select $j$ uniformly at random $i$'s neighbors;
    Generate a random number $q \sim U(0,1)$
  **while** $(q > 1/d_j^\alpha)$
  Return $j$;

As $d_j$ represents the node degree in a connected graph, so $d_j \geq 1$ always holds. So, $1 \geq \frac{1}{d_j^\alpha} \geq \frac{1}{d_j}$ always holds for any positive $\alpha \in (0,1)$. Therefore, when we use algorithm 2 to collect sample, the value of $1 - \tilde{p}_{ii}$ is larger than that of Algorithm 1, but the strength of correcting large deviation is smaller than that of Algorithm 1. Similar with the proof of stationary distribution of Algorithm 1, using $\frac{1}{d_j^\alpha}$ to filter the high-degree nodes can also construct a Markov chain, whose transition matrix is

$$P_{ij} = \begin{cases} \frac{\tilde{p}_{ij}}{1 - \tilde{p}_{ii}}, & if\ v \in N(v) \\ 0, & otherwise \end{cases}$$

where $\tilde{p}_{ij}$ is defined as

$$\tilde{p}_{ij} = \begin{cases} \dfrac{1}{d_i d_j^\alpha}, & if\ j \in N(i) \\ 1 - \sum_{k \in N(i)} \tilde{p}_{ik}, & if\ i = j \\ 0, & otherwise \end{cases}$$

**Theorem 4:** After extended IDRW converges, for any node $i \in V$, the stationary distribution of $i$ being visited $\pi_i = Z \times d_i^{1-\alpha} \times (1 - \tilde{p}_{ii})$, where $Z$ is a normalization constant.

***Proof:*** After the walker reaches equilibrium, the Markov chain constructed by IDRW have the time reversibility. So according to Proposition 1.1 in (Sigman, 2009), we only need to show that the following equation has a unique solution.

$$\pi_i P_{ij} = \pi_j P_{ji}$$

Based on the definition of EIDRW transition matrix, we have

$$\frac{\pi_i}{\pi_j} = \frac{P_{ji}}{P_{ji}} = \frac{\dfrac{1}{d_i d_j^\alpha}(1 - \tilde{p}_{ii})}{\dfrac{1}{d_j d_i^\alpha}(1 - \tilde{p}_{jj})} = \frac{d_i^{1-\alpha}(1 - \tilde{p}_{ii})}{d_j^{1-\alpha}(1 - \tilde{p}_{jj})}.$$

As $d_i^{1-\alpha}(1 - \tilde{p}_{ii})$ and $d_j^{1-\alpha}(1 - \tilde{p}_{jj})$ are fixed for a given graph, so $\dfrac{\pi_i}{\pi_j}$ have a fixed ratio for a given graph. The value of $\pi_i$ is also constricted by the property of a probability density function $\sum \pi_i = 1$, so there is a unique solution for all $\pi_i$, thus for the distribution $\pi$. The stationary distribution of $i$ can be derived as

$$\pi_i = Z \times d_i^{1-\alpha} \times (1 - \tilde{p}_{ii})$$

where $Z$ is the normalization constant. ∎

Similarly, an asymptotically unbiased estimator of extended IDRW is

$$\hat{\theta} = \frac{\sum_{i \in S} w_i f(i)}{\sum_{i \in S} w_i}$$

where $w_i = \dfrac{1}{d_i^{1-\alpha}(1-\tilde{p}_{ii})}$, and $S$ represents the set of sampled nodes.

Note that when $\alpha = 0$, extended IDRW would collapse to SRW. As $\alpha$ is approaching infinity, the limit of the stationary distribution is uniform. Therefore, through adjusting the value of $\alpha$, we can balance the tradeoff between the drawbacks of uniform walk and simple random walk. We would also test the effect of $\alpha$ in the Section 5.

## 5. Numerical Simulation

In this section, we conduct numerical simulation on various real-world network datasets, so that we evaluate the effectiveness and efficiency of IDRW. Experiments results show that

IDRW produces a lower estimation error under the same query cost, thus when we set an error threshold, the query cost of IDRW is smaller than other algorithms. We also reveal fundamental understandings on why IDRW has better performance.

## 5.1 Experimental Setup

We conduct experiments on the datasets released before (Brin and Page, 1998; Leskovec et al., 2008; Rozemberczki et al., 2021; Yang and Leskovec, 2012), including 4 different datasets with different scales. We present basic information of the datasets in Table 1. Note that IDRW is a sampling method that only has "walking" mode, i.e., it can only walk to the neighbor of current node, thus the directed graphs would be transferred into undirected graph, and processed by preserving the largest connected component only. This is a commonly sued method in previous research (Chen et al., 2016; Li et al., 2019; Mohaisen et al., 2012; Nazi et al., 2015).

We compare IDRW with 2 other typical random walk sampling methods: 1) )Non-backtracking Random Walk (NBRW) (Lee et al., 2012) and 2) classical Metropolis-Hasting Random Walk (MHRW). All algorithms are implemented in R, and we tested them on a computer with 1.6 GHz Dual-Core Intel Core i5.

*Table 1 Basic information of datasets*

| Dataset | # of nodes | # of edges | Average Degree |
|---|---|---|---|
| Ca-GrQc | 4158 | 13422 | 6.46 |
| Slashdot | 70999 | 365572 | 10.30 |
| Crocodile | 11631 | 170773 | 29.37 |
| Lastfm | 7624 | 27806 | 7.29 |

## 5.2 Performance metrics

As our target characteristic function is the average degree, the estimation error of average degree we adopt is the relative error.

$$\text{Relative Error} \triangleq |X - \hat{X}|/X$$

where $\hat{X}$ and $X$ are the estimate and the true value of average degree, respectively.

We follow the concept of query cost as the total number of unique queries in a sampling task and do not wait the random walks to get stationary here. For example, if we get a set of sample node from a graph, which is {a, b, c, a, c}, the query cost of the sample is 3. This is a reasonable cost metric and it is also widely used to evaluate the efficiency of sampling algorithms. The reason y we consider the only unique queries is that in the context of online network, the query cost would not be used if the node has been visited before, as the information of visited nodes is stored.

**5.3 Experimental Results**

We present two experiment results in Section 5.3, including the comparisons between 3 different algorithms and the tendency of errors with an increasing parameter $\alpha$.

**Exp-1: Query cost vs. Relative error**

Fig. 2 shows the relationship between query cost and the relative error. As the query cost increases, the relative error decreases, which is consistent with the law of large number (LNN). The black line (IDRW) always has a lower relative error in estimating average degree of graphs under the same query cost, especially when we draw a small proportion of nodes as our sample, IDRW performs much better than the other two algorithms. Fig. 2 shows that the difference in relative error when sample size is between 100 to 300 is much larger than that with sample size over 600. One explanation for this occurrence is that when the sample size increases, the estimated values are closer to the true values. In particular, for the MHRW algorithm, setting the same query cost causes the MH algorithm to collect more samples, thus making the MH random walker reach a steady state. Therefore, one recommended application of IDRW is when the online platforms restricted the query cost, IDRW can perform better than other algorithms.

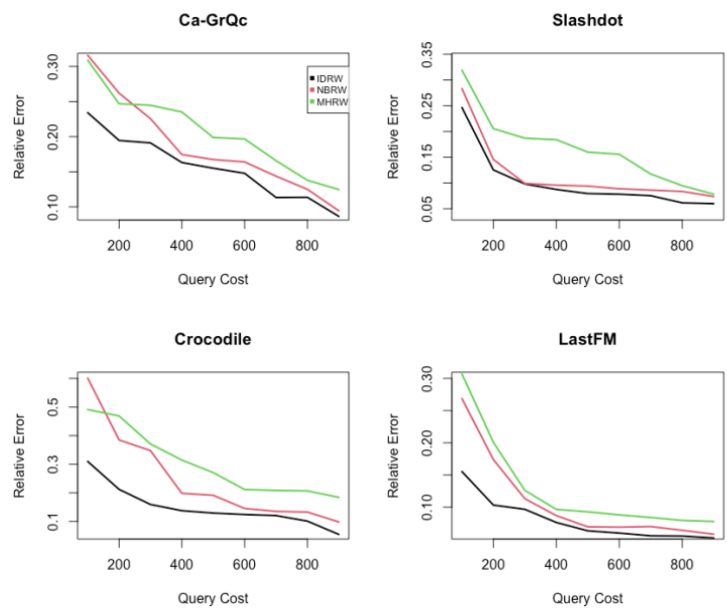

*Figure 2 Comparison between different algorithms*

**Exp-2: the effect of $\alpha$**

We change the value of $\alpha$ from 0 to 1 with an interval of 0.1. The results with varying $\alpha$ is in the Fig. 3. Although extended IDRW with an increasing $\alpha$ on different datasets shows different behaviors, we can see that when we keep $\alpha$ in a low level, the error is relatively smaller than that of an $\alpha$ close with 1. According to the mathematical analysis of our newly proposed method, the stationary distribution is proportional to the value of $d_i^{1-\alpha} \times (1 - \tilde{p}_{ii})$. If we go a further step, the probability of node $i$ being sampled is proportional to $d_i^{-\alpha} \sum_{j \in N(i)} d_j^{-\alpha}$. When $\alpha = 1$, the probability of node $i$ being chosen is the inverse of harmonic average of $i$'s neighbor degrees. When we use IDRW with $\alpha = 1$ to collect data, the higher degree the node has, the more overcorrection it will receive. Otherwise, low-degree nodes usually do not change very much due to both low degree and possibly low local density. When we lower down the power of degrees, the overcorrection is not that much thus the stationary distribution is more similar to the uniform distribution. Hereby, we also give a recommended range of application $\alpha \in [0.1, 0.4]$. For many real-world data, parameter between this range perform better than the unadjusted inverse degree.

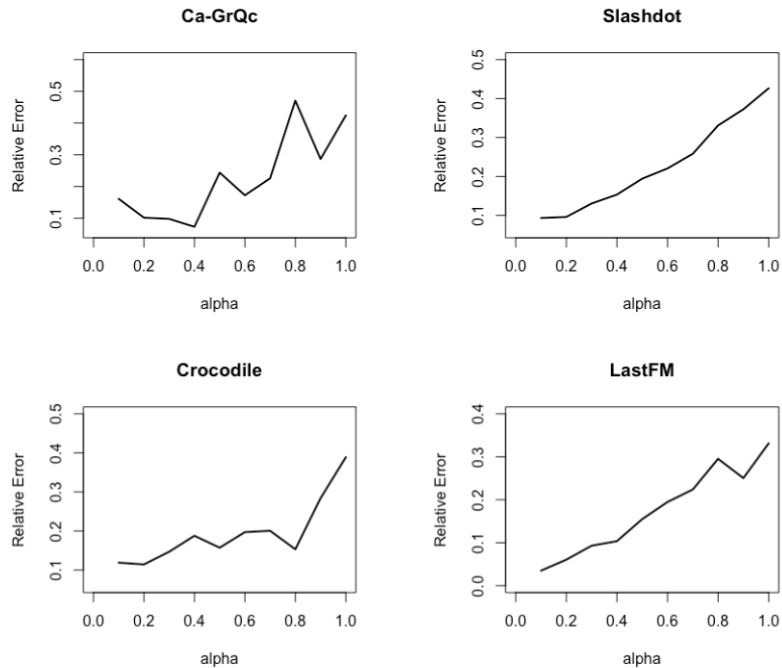

Figure 3 alpha vs. relative error

## 6. Conclusion

In this paper, we proposed a framework that can modify uniform walking in SRW to weighted walking, correcting the bias towards higher degree nodes. Intuitively, we devise the walking strategies to lower down the inclusion probability of high-degree nodes while higher up the inclusion probability of low-degree nodes. We also consider the rejection problem in sampling procedure, so we add a parameter to adjust the power of inversed degree value. With an adjustable parameter $\alpha$, we can choose different $\alpha$ as needed to achieve optimal results. We also conduct extensive experiments in this paper, not only investigate the effect of query cost on error, but also how $\alpha$ affect the error. Our framework provides a flexible series of methods that can be applied to many OSN sampling tasks by adjusting parameters.